%% file: teq.tex
\documentclass[11pt]{article}

\usepackage[margins,times,theorems,greekit]{pamas}
\usepackage{ifthen,mathrsfs}
\usepackage{pgflibraryshapes}
\usepackage{algorithm,algorithmic}

\newcommand{\romanenumi}{\renewcommand{\theenumi}{({\it\roman{enumi}})}
    \renewcommand{\labelenumi}{\theenumi}}

\newcommand{\set}[1]{\{#1\}}

\newcommand{\dom}{\overline{D}}
\newcommand{\tc}{\mathit{TC}}

\newcommand{\teq}{TEQ\xspace}
\newcommand{\teqset}[1][]{\ifthenelse{\equal{#1}{}}{\text{\it TEQ}}{\text{\it TEQ}(#1)}}
\newcommand{\teqrel}[1][]{\ifthenelse{\equal{#1}{}}{\boldsymbol\rightarrow}{\teqrel_{#1}}}
\newcommand{\weg}[1]{}

\title{Recognizing Members of the Tournament Equilibrium Set is NP-hard
}
\author{%
	Felix Brandt \qquad Felix Fischer \qquad Paul Harrenstein \\
	Institut f\"ur Informatik, Universit\"at M\"unchen, Germany \\
	\texttt{\small\{brandtf,fischerf,harrenst\}@tcs.ifi.lmu.de}}
\date{}

\begin{document}
\maketitle

\begin{abstract}
  A recurring theme in the mathematical social sciences is how to select the ``most desirable'' elements given a binary dominance relation on a set of alternatives. Schwartz's tournament equilibrium set (\teq) ranks among the most intriguing, but also among the most enigmatic, tournament solutions that have been proposed so far in this context. Due to its unwieldy recursive definition, little is known about \teq.  In particular, its monotonicity remains an open problem up to date. Yet, if \teq were to satisfy monotonicity, it would be a very attractive tournament solution concept refining both the Banks set and Dutta's minimal covering set. We show that the problem of deciding whether a given alternative is contained in \teq is NP-hard.
\end{abstract}

\section{Introduction}

The central problem of the literature on tournament solutions is as appealing as it is simple: Given an irreflexive, asymmetric, and complete binary relation over a set, find the most attractive elements of this set. As the standard notion of maximality is not well-defined in the presence of cycles, numerous alternative solution concepts have been devised and axiomatized \citep[see, \eg][]{Moul86a,Lasl97a}. In social choice theory, the base relation, which we call dominance relation, is commonly defined via pairwise majority voting, and many well-known tournament solutions yield attractive social choice correspondences.


Over the years, the computational complexity of almost all common solution concepts has been completely characterized~\citep[see, \eg][]{Woeg03a,Coni06a,BFH07b,BrFi07a}.  One notable exception is the tournament equilibrium set (\teq) proposed by \citet{Schw90a}.
Due to its unwieldy recursive definition, little is known about \teq \citep{Dutt90a,LLL93a}.  However, if a certain technical conjecture stated almost two decades ago turned out to be true, it would constitute one of the most attractive tournament solutions, refining both the minimal covering set and the Banks set \citep{Lasl97a,LLL93a}. \citeauthor{Lasl97a} states that ``Unfortunately, no algorithm has yet been published for finding the Minimal Covering set or the tournament equilibrium set of large tournaments. For tournaments of order 10 or more, it is almost impossible to find (in the general case) these sets at hand''~\citep[p.~8]{Lasl97a}.
While it has recently been shown that computing the minimal covering set is feasible in polynomial time \citep{BrFi07a}, it turns out that this is not the case for \teq unless P equals NP. We prove this by first providing an alternative hardness proof for the Banks set which is then modified so as to apply to \teq as well.  In contrast to the Banks set, there is no obvious reason to suppose that the \teq membership problem is in NP; it may very well be even harder.

\section{Preliminaries}

A \emph{tournament $T$} is a pair $(A,\succ)$, where $A$ is a finite set of \emph{alternatives} and $\succ$ an irreflexive, anti-symmetric, and complete binary relation on $A$, also referred to as the \emph{dominance relation}.  Intuitively, $a\succ b$ signifies that alternative $a$ beats $b$ in a pairwise comparison. We write $\mathcal T$ for the class of all tournaments and we have $\mathcal T(A)$ denote the set of all tournaments on a fixed set~$A$ of alternatives. If $T$ is a tournament on~$A$, then every subset $X$ of $A$ induces a tournament $T|_X$ defined as $(X,\mathord{\succ}\mathord{|}_X)$, where $\mathord{\succ}\mathord{|}_X=\set{(x,y)\in X\times X\colon x\succ y}$.

As the dominance relation is not assumed to be transitive in general, it may contain cycles. Moreover, there need not be a so-called \emph{Condorcet winner}, \ie an alternative that dominates all others. This makes that the usual notions of maximum and maximal elements are no longer feasible in this context. Other concepts have been suggested to take over the role of singling out those alternatives that can somehow be considered the ``winners'' of the tournament. Formally, a \emph{tournament solution $S$} is defined as a function that associates with each tournament $T$ on $A$ a subset~$S(T)$ of $A$. The definition of a tournament solution commonly includes the requirement that $S(T)$ be non-empty if $T$ is defined on a non-empty set of alternatives and that it select the Condorcet winner if there is one \citep[p.37]{Lasl97a}.
For~$X$ a subset of~$A$, we also write~$S(X)$ for the more cumbersome~$S(T|_{X})$, provided that the tournament~$T$ is known from the context. In this paper we will be concerned with two particular tournament solutions, the Banks set and Schwartz's tournament equilibrium set (\teq). For a proper definition, however, we need some auxiliary notions and notations.

Let~$R$ be a binary relation on a set $A$. We write~$R^*$ for the transitive reflexive closure of~$R$. By the \emph{top cycle~$\tc_A(R)$} we understand the maximal elements of the asymmetric part of~$R^*$. 
A subset~$X$ of~$A$ is said to be \emph{transitive} if~$R$ is transitive on~$X$. 
For $X\subseteq Y\subseteq A$, $X$ is called \emph{maximal in~$Y$} if no proper superset of~$X$ in~$Y$ is transitive, \ie if there is no transitive $Z\subseteq Y$ with $X\subset Z$. Clearly, every transitive set is contained in a maximal transitive set. Given a set $Z=\set{Z_i}_{i\in I}$ of pairwise disjoint subsets of~$A$, a subset~$X$ of~$A$ will be called a \emph{choice set for~$Z$} if it contains precisely one element from each subset $Z_i\in Z$.

In tournaments, maximal transitive sets are also referred to as Banks trajectories. The \emph{Banks set~$BA(T)$} of a tournament~$T$ then collects the maximal elements of the Banks trajectories.
\begin{definition}[Banks set] Let~$T$ be a tournament on~$A$.
 An alternative $a\in A$ is in the \emph{Banks set~$BA(T)$} of~$T$ if~$a$ is a maximal element of some maximal transitive set in~$T$.
\end{definition}

The \emph{tournament equilibrium set} of a tournament~$T$ on~$A$ is defined as the top cycle of a particular subrelation of the dominance relation, the \teq relation.
The underlying idea is that an alternative is only ``properly'' dominated, \ie dominated according to the subrelation, if it is dominated by an element that is selected by some tournament solution concept~$S$. To make this idea precise, for~$X\subseteq A$, we write $\dom_X(a)=\{\,b\in X\colon b\succ a\,\}$ for the \emph{dominators} of~$a$ in~$X$, omitting the subscript when $X=A$. Thus, for each alternative~$a$ one examines the dominator set~$\dom(a)$, and solves the subtournament $T|_{\dom(a)}$ by means of the solution~$S$. In the subrelation~$a$ is then only dominated by the alternatives in $S(\dom(a))$. This of course, still leaves open the question as to the choice of the solution concept~$S$. Now, in the case of $\teqset$, $S$ is taken to be $\teqset$ itself! The reason why this is a proper recursive definition is that if $a\in X$ for some $X\subseteq A$ the set~$\dom_X(a)$ of dominators of~$a$ in~$X$ is a proper subset of~$X$. \Ie in order to determine the \teq relation in a particular subtournament~$T$, one has to calculate the \teq of a proper subtournament of~$T$.

\begin{definition}[Tournament equilibrium set] Let $T$ be a tournament on $A$. For each subset $X$ of $A$ we define the \emph{tournament equilibrium set} $\teqset[X]$ for $X$ as follows:
 	\[
		\teqset[X]=\tc_X({\teqrel[X]}),
	\]
	where $\teqrel[X]$ is defined as the binary relation on $X$ such that for all $x,y\in X$,
	\[
	x\teqrel[X]y\text{ if and only if }x\in\teqset[{\dom_X(y)}].
	\]
\end{definition}
Observe that the \teq relation~$\teqrel[X]$ is invariably a subset of the dominance relation~$\succ$ and that if $\dom_X(x)\not=\emptyset$, then there is some $y\in\dom_X(x)$ with $y\teqrel[X]x$.

It can easily be established that the Banks set and \teq both select the Condorcet winner in a tournament if there is one. Moreover, in a cyclic tournament on three alternatives, the Banks set and \teq both consist of all alternatives. Yet, the Banks set and \teq do not coincide for all tournaments. For an example consider the tournament depicted in \figref{fig:tournament_example}. 
First we calculate the \teq relation~$\teqrel$. Observe that $\dom(a)=\set{c}$. Hence,~$c$ is the Condorcet winner in $\dom(a)$ and we have $\teqset[\dom(a)]=\set c$ and $c\teqrel a$. For alternative~$b$ we have
$\dom(b)=\set{a,e}$. Since $a\succ e$, alternative~$a$ is the Condorcet winner in~$\dom(b)$, and we may conclude that $\teqset[\dom(b)]=\set a$. Hence, $a\teqrel b$ whereas $e\not\teqrel b$. In an analogous fashion we find for all $x\in\set{a,b,c,d,e}$ that $x\teqrel c$ if and only if $x=b$ as well as that $x\teqrel d$ if and only if $x=a$. For alternative~$e$, however, 
$\dom(e)=\set{a,c,d}$. Since $a\succ d\succ c\succ a$, a three-cycle, we have that $\teqset[\dom(e)]=\set{a,c,d}$ and hence $a\teqrel e$, $c\teqrel e$ as well as $d\teqrel e$. The top cycle of the relation~$\teqrel$ thus found can then be seen to coincide with the set~$\set{a,b,c}$, which then also constitutes the \teq of this tournament. By contrast, the Banks set in this example consists of the four elements~$a$,~$b$,~$c$ and~$d$. Maximal transitive sets of which $a$, $b$, $c$ and $d$ are the maximal elements are, \eg
$\set{a,b,d}$, $\set{b,c}$, $\set{c,a,e}$ and $\set{d,c,e}$, respectively. Alternative~$e$ is not included in the Banks set. The only transitive subsets of which~$e$ is the maximal element are $\set e$ and $\set{e,b}$. However, both these sets are included in maximal transitive sets of which $e$ is not the maximal element, \eg the set $\set{a,e,b}$. Thus \teq and the Banks set may differ.  However, the former is known to always be included in the latter.
\begin{figure}[tb]
  \centering
  \input{Tournament_Example}
  \caption{Example due to \citealp{Schw90a}, where $BA(T)=\set{a,b,c,d}$ and $\teqset[T]=\set{a,b,c}$. The relation~$\teqrel$ is indicated by thick edges.}
  \label{fig:tournament_example}
\end{figure}

\begin{proposition}[\citealp{Schw90a}]\label{prop:TEQinBanks}
 Let $T=(A,\succ)$ be a tournament. Then, $\teqset[T]\subseteq BA(T)$.
\end{proposition}
\begin{proof}
 We prove by structural induction on~$X$ that~$\teqset[X]\subseteq BA(X)$ for all subsets~$X$ of~$A$. The case $X=\emptyset$ is trivial, as then $\teqset[X]= BA(X)=\emptyset$. So, assume that $\teqset[X']\subseteq BA(X')$, for all $X'\subsetneq X$. We prove that $\teqset[X]\subseteq BA(X)$ as well. To this end, consider an arbitrary $a\in \teqset[X]$. Either $\dom_X(a)=\emptyset$ or $\dom_X(a)\neq\emptyset$. In the former case,~$a$ is the Condorcet winner in~$X$ and therefore $a\in BA(X)$. In the latter case, $x\teqrel[X]a$ for some $x\in X$. Having assumed that $a\in\teqset[X]$, \ie $a\in\tc(\teqrel[X])$, there is also an $x'\in X$ with $a\teqrel[X]x'$. Accordingly, $a\in\teqset[{\dom_X(x')}]$. By the induction hypothesis, also $a\in BA(\dom_X(x'))$. Therefore, there is some maximal transitive set~$Y$ in $\dom_X(x')$ of which~$a$ is the maximal element. Then, $Y\cup\set{x'}$ is a transitive set in $X$. Now let~$Y'\subseteq X$ be a maximal transitive set in~$X$ containing~$Y\cup\set{x'}$ with~$a'$ as maximal element. Observe that $a'\in BA(X)$. Then, $a'\succ x'$ and so $a'\in\dom_X(x')$. Now consider $Y'\cap\dom_X(x')$. Clearly, $Y\cap\dom_X(x')$ is a transitive set in $\dom_X(x')$ which contains~$a'$ as its maximal element. Moreover, $Y\subseteq Y'\cap\dom_X(x')$. By maximality of $Y$ it then follows that $Y=Y'\cap\dom_X(x')$ and that $a=a'$. We may conclude that $a\in BA(X)$.\qed
\end{proof}

 In the remainder of this paper, we assume the reader to be familiar with the well-known complexity classes~P and~NP and the notion of polynomial-time reducibility~\citep[see, \eg][]{Papa94a}.

\section{A Heuristic for Computing the Tournament Equilibrium Set}

It is not very hard to see that the naive algorithm for computing \teq, which simply recursively computes~$\teqrel$, requires exponential time in the worst case. Running time can be greatly reduced by using a heuristic that relies on the conjecture that the top cycle of any \teq relation consists of only one strongly connected component. This conjecture was already made by \citet{Schw90a} and has later been shown to be equivalent to the (conjectured) monotonicity of \teq~\citep{LLL93a}. 
Algorithm~\ref{alg:teq} computes \teq by starting with the set $B$ of all alternatives that have dominator sets of minimal size. These alternatives are likely to be included in \teq and the small size of their dominator sets speeds up the determination of their \teq-dominators. In the following, all alternatives that \teq-dominate any alternative in $B$ are iteratively added to $B$ until no more such alternatives can be found, in which case the algorithm returns the top cycle of $B$.
Clearly, the \emph{worst-case} running time of this algorithm is still exponential, and it will be shown in the remainder of the paper that this has to be the case for every algorithm computing \teq unless $P=NP$.

\begin{algorithm}[tb]
\begin{algorithmic}
	\STATE \textbf{procedure} $\texttt{TEQ}(X)$
	\STATE $R\leftarrow \emptyset$
	\STATE $B\leftarrow C\leftarrow \arg \min_{a\in X} |\dom(a)|$	
	\LOOP
	\STATE $R\leftarrow R\cup \{ (b,a) \colon a\in C \wedge b\in \texttt{TEQ}(\dom(a))\}$
	\STATE $D\leftarrow \bigcup_{a\in C} \texttt{TEQ}(\dom(a))$
	\STATE \textbf{if} $D\subseteq B$ \textbf{then return} $\tc_B(R)$ \textbf{end if}
	\STATE $C\leftarrow D$
	\STATE $B\leftarrow B\cup C$
	\ENDLOOP
\end{algorithmic}
\caption{Tournament Equilibrium Set}
\label{alg:teq}
\end{algorithm}

\section{An Alternative NP-Hardness Proof for Membership in the Banks Set}

The problem of deciding whether a particular alternative is included in the Banks set of a given tournament $T$ is known to be NP-complete. This was first demonstrated by \cite{Woeg03a} by means of a reduction from graph three-colorability. Here we will give an alternative proof of this result. Our proof works by a reduction from $\mathit{3SAT}$, the NP-complete satisfiability problem for Boolean formulas in conjunctive normal form with exactly three literals per clause~\citep[see, \eg][]{Papa94a}. The construction used in this paper is arguably simpler than Woeginger's. Moreover, a much similar construction will be used in the next section to prove NP-hardness of the analogous decision problem for \teq. The tournaments used in these reductions will both be taken from a special class $\mathcal{T}^*$, which we introduce next.

\begin{definition}[The class $\mathcal T^\ast$]\label{def:tstar}
	A tournament $(A,\succ)$ is in the class $\mathcal T^\ast$ if it satisfies the following properties. There is some odd integer $n\ge 1$, the \emph{size} of the tournament, such that $A=C\cup U_1\cup\cdots\cup U_n$, where $C,U_1,\ldots, U_n$ are pairwise disjoint and $C=\set{c_0,\ldots,c_n}$. Each $U_i$ is a singleton if $i$ is even, and $U_i=\set{u_i^1,u_i^2,u_i^3}$ if $i$ is odd.
	The complete and asymmetric dominance relation~$\succ$ is such that the following five properties hold 	for all $c_i\in C_i$, $c_j\in C_j$, $u_i\in U_i$, $u_j\in U_j$ ($0\le i,j\le n$):
	\begin{enumerate}\romanenumi
	 \item\label{itm:TstarCnodes} 	$c_i\succ c_j$,\quad  if $i<j$,
	 \item\label{itm:tstarCUnodesi}	$u_i\succ c_j$,\quad if $i=j$,
	 \item\label{itm:TstarCUnodesii}  $c_j\succ u_i$,\quad if $i\not=j$,
    	 \item\label{itm:separating}   $u_i\succ u_j$,\quad if $i<j$ and at least one of $i$ and $j$ is even\quad (\ie $u_i$ or $u_j$ is a separating node),
	 \item\label{itm:tstarxbarx}	$u^k_i\succ u^l_i$,\quad if $i$ is odd and $k\equiv l-1\pmod{3}$ \quad (\ie $u_i^1\succ u^2_i\succ u^3_i\succ u^1_i$).
	\end{enumerate}

We also refer to $c_0$ by $d$, for ``decision node'' and to $\bigcup_{1\le i\le n}U_n$ by $U$. For $i=2k$, we have as a notational convention $U_i=Y_k=\set{y_k}$ and set $Y=\bigcup_{1\le 2k\le n}Y_k$. These nodes are called \emph{separating nodes}.
\end{definition}
Observe that this definition fixes the dominance relation between any two alternatives except for some pairs of alternatives that are both in~$U$. 

As a next step in the argument, we associate with each instance of~$\mathit{3SAT}$ a tournament in the class~$\mathcal T^*$. An instance of~$\mathit{3SAT}$ is given by a formula~$\varphi$ in \emph{$3$-conjunctive normal form~($\text{3CNF}$)}, \ie $\varphi=(x^1_1\vee x^2_1\vee x^3_1)\wedge\cdots\wedge(x^1_m\vee x^2_m\vee x^3_m)$, where each $x\in\set{x^1_i,x^2_i,x^3_i\colon 1\le i\le m}$  is a literal. For each clause $x^1_i\vee x^2_i\vee x^3_i$ we assume~$x^1_i$, $x^2_i$ and~$x^3_i$ to be distinct literals. We moreover assume the literals to be indexed and by $X_i$ we denote the set $\set{x_i^1,x_i^2,x_i^3}$.
For literals~$x$ we have $\bar x=\neg p$ if $x=p$, and $\bar x=p$ if $x=\neg p$, where~$p$ is some propositional variable.
We may also assume that if $x$ and $y$ are literals in the same clause, then $x\not=\bar y$. We say a $\text{3CNF}$ $\varphi=(x^1_1\vee x^2_1\vee x^3_1)\wedge\cdots\wedge(x^1_m\vee x^2_m\vee x^3_m)$ is \emph{satisfiable} if there is a choice set $V$ for $\set{X_i}_{1\le i\le m}$ such that $v'=\bar v$ for no $v,v'\in V$. Next we define for each $\mathit{3SAT}$ formula~$\varphi$ the tournament~$T^{BA}_\varphi$.

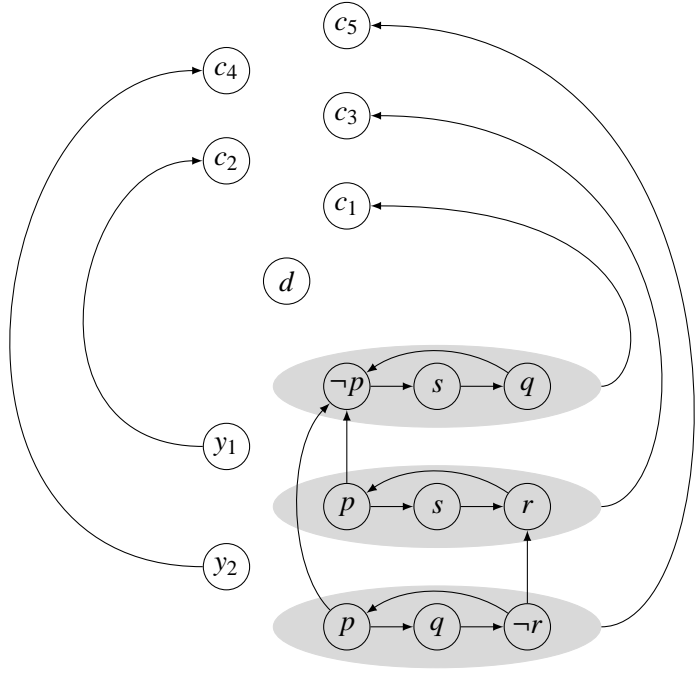
\begin{figure}[tb]
  \centering
  \input{Banks_construction_form}
  \caption{Tournament $T^{BA}_{\varphi}$ for the $3$-CNF formula $\varphi=(\neg p\vee s\vee q)\wedge(p\vee s\vee r)\wedge(p\vee q\vee \neg r)$.}
  \label{fig:banks}
\end{figure}

\begin{definition}[Banks construction]\label{def:Banks_construction}
	Let $\varphi$ be a $3$CNF $(x^1_1\vee x^2_1\vee x^3_1)\wedge\cdots\wedge(x^1_m\vee x^2_m\vee x^3_m)$.  Define $T^{BA}_{\varphi}=(C\cup U,\succ)$ as the tournament in the class $\mathcal T^\ast$ of size $2m-1$ such that for all $1\le j< 2m$,
	\[
        U_j=
        \begin{cases}   X_i &\text{if $j=2i-1$,}\\
                        \set{y_i} &\text{if $j=2i$}
        \end{cases}
        \]
        and such that for all $x\in X_i$ and $x'\in X_j$ ($1\le i,j\le m$),
        \[
            \text{$x\succ x'$ \quad {if both $j<i$ and $x'=\bar x$ or both $i<j$ and $x'\not= \bar x$.}}
	\]
Observe that in conjunction with the other requirements on the dominance relation of a tournament in $\mathcal T^\ast$, this completely fixes the dominance relation~$\succ$ of $T^{BA}_\varphi$.
\end{definition}
An example for such a tournament is shown in \figref{fig:banks}.
We are now in a position to prove NP-completeness of deciding whether a particular alternative is in the Banks set.

\begin{theorem}
	\label{thm:banks}
	The problem of deciding whether a particular alternative is in the Banks set of a tournament is NP-complete.
\end{theorem}
\begin{proof}
\emph{Membership} in NP is obvious. For a fixed alternative~$d$, we can simply guess a transitive subset of alternatives~$V$ with~$d$ as maximal element and verify that~$V$ is also maximal w.r.t.\@ set inclusion.

 For NP-hardness, we show that $T_\varphi^{BA}$ contains a maximal transitive set with maximal element~$d$ if and only if $\varphi$ is satisfiable. First observe that~$V$ is a maximal transitive subset with maximal element~$d$ in~$T_\varphi^{BA}$ only if 
 \begin{enumerate}
   \romanenumi
   \item\label{itm:everylevel} for all $1\le i <2m$ there is a $u\in U_i$ such that $u\in V$, and
   \item\label{itm:nocycle} there are no $1\le i<j <2m$, $u\in U_i$, $u'\in U_j$ with $u,u'\in V$ such that $u_j\succ u_i$.
 \end{enumerate}
Regarding \ref{itm:everylevel}, if there is an $1\le i <2m$ such that no element of~$U_i$ is contained in~$V$, we can always add~$c_i$ to~$V$ in order to obtain a larger transitive set. If \ref{itm:nocycle} were not to hold, both~$i$ and~$j$ have to be odd for~$u_j$ to dominate $u_i$. However, in light of \ref{itm:everylevel}, there has to be~$k$ with $i<k<j$ and $u''\in U_k$ such that $u''\in V$. It follows that $V$ is not transitive because~$u$,~$u''$, and~$u'$ form a cycle. If there is maximal transitive set $V$ with maximal element $d$ complying with both \ref{itm:everylevel} and \ref{itm:nocycle}, a satisfying assignment of $\varphi$ can be obtained by letting all literals contained in $X\cap V$ be true.
 
 For the opposite direction, assume that $\varphi$ is satisfiable. Then there is a choice set $W$ for $\set{X_i}_{1\le i\le m}$ such that $x'=\bar x$ for no $x,x'\in W$. Obviously $V=W\cup\set{y_1,\ldots,y_{m-1}}\cup \{d\}$ does not contain any cycles and thus is transitive with maximal element~$d$. In order to obtain a larger transitive set with a different maximal element, we need to add~$c_i$ for some $1\le i\le m$ to~$V$. However, $V\cup \{c_i\}$ always contains a cycle consisting of $c_i$, $d$, and $u$ for some $u\in U_i$, contradicting the transitivity of $V\cup \{c_i\}$. We have thus shown that~$d$ is the maximal element of some maximal transitive set  in $T_\varphi^{BA}$ containing~$V$ as a subset. \qed
\end{proof}

\section{NP-hardness of Membership in the Tournament Equilibrium Set}
\label{sec:teqhardness}

As the main result of this paper, we demonstrate that the problem of deciding whether a particular alternative is in the \teq of a tournament is NP-hard.  To this end, we refine the construction used in the previous section to prove NP-completeness of membership in the Banks set.

\begin{figure}[tb]
  \centering
  \input{TEQ_construction_form}
  \caption{Tournament $T^{\teqset}_{\varphi}$  for the $3$-CNF formula $\varphi=(\neg p\vee s\vee q)\wedge(p\vee s\vee r)\wedge(p\vee q\vee \neg r)$.
Omitted edges are again assumed to point downwards.}
  \label{fig:teq}
\end{figure}
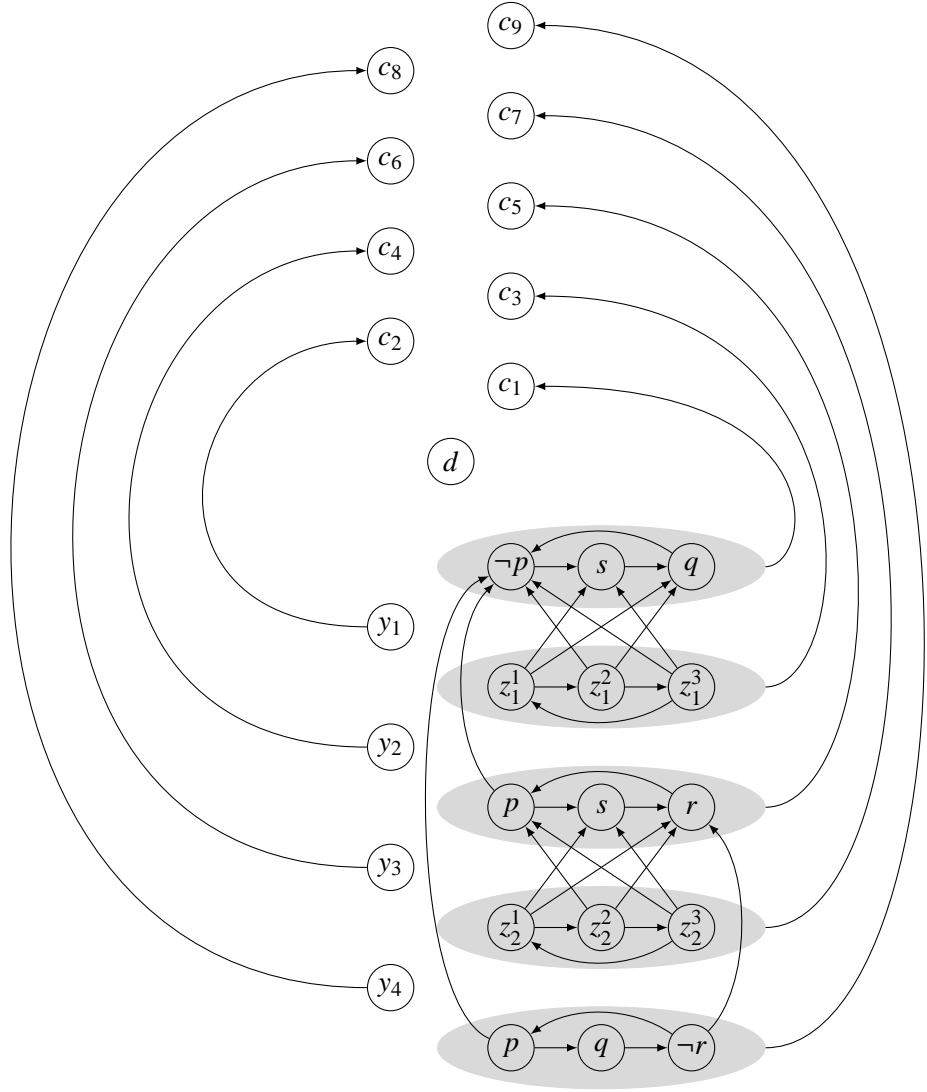

\begin{definition}\label{def:TEQ_construction}
	Let $\varphi$ a $3$-conjunctive normal form $(x^1_1\vee x^2_1\vee x^3_1)\wedge\cdots\wedge(x^1_m\vee x^2_m\vee x^3_m)$. Further for each $1\le i<m$, let there be a set $Z_i=\set{z_i^1,z_i^2,z_i^3}$.
	 Define $T^{\teqset}_{\varphi}$  as the tournament $(A,\succ)$ in $\mathcal{T}^*$ of size $4n-3$ such that
	$A=C\cup U_1\cup\cdots\cup U_{4n-3}$ and for all $1\le i\le m$,
	\[
		U_j=
		\begin{cases}
		 X_i	& 	\text{if $j=4i-3$,}\\
		 Z_i	&	\text{if $j=4i-1$,}\\
		 \set{y_i}	&	\text{otherwise.}
		\end{cases}
	\]
	As in the Banks construction, we let for all $x\in X_i$ and $x'\in X_j$ ($1\le i,j\le m$)
        \[
            \text{$x\succ x'$ \quad {if both $j<i$ and $x'=\bar x$ or both $i<j$ and $x'\not= \bar x$.}}
	\]
	Finally, for all $1\le i,j\le m$, $x_i^k\in X_i$ and $z_j^l\in Z_j$,
	\[
		x^k_i\succ z^l_j\text{ if and only if $i<j$ or both $i=j$ and $k=l$.}
	\]
%
%
\end{definition}
An example for such a tournament is shown in \figref{fig:teq}.

We now proceed to show that a~$\mathit{3SAT}$ formula~$\varphi$ is satisfiable if and only if the decision node~$d$ is in the tournament equilibrium set of~$T^{\teqset}_\varphi$. 

\begin{lemma}\label{lemma:CteqdomX}
	Let $T=(C\cup U,\succ)$ be a tournament in $\mathcal{T}^*$ and let $B\subseteq C\cup U$ such that $d\in B$.  Then, for each $u\in U\cap B$ there exists $c\in C\cap B$ such that $c\teqrel[B]^*u$
\end{lemma}

\begin{proof}
	Let $c_i\in C\cap B$ be such that $\dom_{B}(c_i)\cap C=\emptyset$, \ie $c_i$ is the alternative in $C$ with the highest index among those included in $B$.  Then,
	\begin{equation}
		\label{lemmaitem:constpropsi}
		\text{$c_i\teqrel[B]c$ for all $c\in B\cap C$ with $c\neq c_i$.} 
	\end{equation}
For this, merely observe that by construction $c_i$ is the Condorcet winner in $\overline D_B(c)$. Hence, $c_i\in\teqset[{\overline D_B(c)}]$ and $c_i\teqrel[B]c$. 

The lemma itself then follows from the stronger claim that for each $u\in U\cap B$ there 
is some $c\in C\cap B$ with both $c\teqrel[B]^*u$ and $c\in\teqset[B]$.
This claim we prove by structural induction on supersets~$B$ of~$\set d$. 

If $B=\set d$, $U\cap B=\emptyset$ and the claim is satisfied trivially. So let $\set d$ be a proper subset of~$B$. Again, if $U\cap B=\emptyset$, the claim holds trivially. So we may assume there be some $u\in U\cap B$. Then, $d\in\dom_B(u)$ by construction of~$T$. If $\dom_B(u)\cap U=\emptyset$, $\dom_B(u)$ is a non-empty subset of~$C\cap B$, and so is $\teqset[\dom_B(u)]$. It follows that for some $c\in\teqset[\dom_B(u)]\cap C$ we have $c\teqrel[B]u$. If, on the other hand, $\dom_B(u)\cap U\neq\emptyset$, the induction hypothesis is applicable and we have $c\in\teqset[\dom_B(u)]$ for some $c\in C\cap B$. Hence, $c\teqrel[B]u$.
With~$u$ having been chosen arbitrarily, we actually have that for all $u\in U\cap B$, there is some $c\in C\cap B$ with $c\teqrel[B]u$. It remains to be shown that there is some $c\in C\cap\teqset(B)$ with $c\teqrel[B]^*u$.

To this end, again consider $c_i\in C\cap B$ such that $\dom_{B}(c_i)\cap C=\emptyset$.  It suffices to show that $c_i\teqrel[B]^*b$ for all $b\in B$, as then both $c_i\in\teqset(B)\cap C$ and $c_i\teqrel[B]^*u$. So, consider an arbitrary $b\in B$. If $b=c_i$, the case is trivial. If $b\in C\cap B$ but $b\neq c_i$, we are done by~(\ref{lemmaitem:constpropsi}). If instead $b\in U\cap B$, then $c\teqrel[B]^*b$ for some $c\in C\cap B$, as we have shown in the first part of the proof.  If $c=c_{i}$, we are done.  Otherwise, we can apply~(\ref{lemmaitem:constpropsi}) to obtain $c_i\teqrel[B]c'\teqrel[B]^*b$ and hence $c_i\teqrel[B]^*b$.
%
%
\qed
\end{proof}

\begin{theorem}
 Deciding whether a particular alternative is in the tournament equilibrium set of a tournament is NP-hard.
\end{theorem}
\begin{proof}
By reduction from $3$-SAT. Consider an arbitrary $3$-DNF $\varphi$ and construct the tournament $T^{\teqset}_\varphi=(C\cup U,\succ)$. This can be done in polynomial time. We show that
\[ 
\text{$\varphi$ is satisfiable if and only if $d\in\teqset[{T^{\teqset}_\varphi}]$. }
\]
For the direction from left to right, observe that by an argument analogous to the proof of Theorem~\ref{thm:banks} it can be shown that~$\varphi$ is satisfiable if and only if $d\in BA(T^{\teqset}_\varphi)$. So assuming that 
$\varphi$ is not satisfiable yields $d\notin BA(T^{\teqset}_\varphi)$.  By the inclusion of \teq in the Banks set (Proposition~\ref{prop:TEQinBanks}), it follows that $d\notin\teqset[{T^{\teqset}_\varphi}]$.

For the opposite direction, assume that $\varphi$ is satisfiable. Then there is a choice set $W$ for $\set{X_i}_{1\le i\le m}$ such that $x'=\bar x$ for no $x,x'\in W$. Obviously $W\cup\set{y_1,\ldots,y_{m-1}}\cup\set{z^j_i\in Z\colon x^j_i\in W}=\set{u_1,\ldots,u_n}$ contains no cycles and thus is transitive. Without loss of generality we may assume that $u_i\in U_i$ for all $1\le i\le n$. For each $1\le i\le n+1$, define a subset $\overline D^{u_1,\ldots,u_n}_{i}$ of alternatives as follows. Set
 $\overline D^{u_1,\ldots,u_n}_{n+1}=A$ and let $\overline D_i^{u_1,\ldots,u_n}$ denote $\bigcap_{i\le j\le {n+1}}\overline D(u_j)$ for each $1\le i\le n$. Hence, $\dom_1^{u_1,\ldots,u_n}\subsetneq\cdots\subsetneq\dom_{n+1}^{u_1,\ldots,u_n}$. In an effort to simplify notation, we write $\teqrel[i]$ and $\overline{D}_i(x)$ for $\teqrel[\overline D^{u_1,\ldots,u_n}_i]$ and $\overline D_{\overline D^{u_1,\ldots,u_n}_i}(x)$, respectively. We will also assume $u_1,\ldots,u_n$ to be fixed and write $\dom_i$ for $\dom_i^{u_1,\ldots,u_n}$. It then suffices to prove that
\begin{equation}
	\label{eqn:dinteq}
	\text{$d\in\teqset[\overline D_{k}]$, for all $1\le k\le n+1$.}
\end{equation}
The theorem then follows as the special case in which $k=n+1$. We first make the following observations concerning the \teq relation $\teqrel[i]$ in each $\overline D_i$, for each $1\le i,j\le n+1$:
\begin{enumerate}
\romanenumi
 	\item\label{itm:2i} 	$u_j\in\overline D_{i}$ if and only if $j<i$,
	\item\label{itm:2half}	$c_j\in\overline D_{i}$   if and only if $j<i$,
	\item\label{itm:2ii}	$c_{i}\teqrel[i+1]c_j$ if $j<i\le n$,
	\item\label{itm:2iii}	$u_i\teqrel[i+1] c_{i}$, if $i\le n$.
\end{enumerate}
	For~\ref{itm:2i}, observe that if $j<i$, $u_j\in\overline D(u_i)$ by transitivity of the set $\set{u_1,\ldots,u_n}$. Hence, $u_j\in\overline D_i$. If on the other hand $j\ge i$, then $u_j\notin\dom(u_j)$ and thus $u_j\notin\dom_i$. For~\ref{itm:2half}, observe that $c_j\in\dom(u_i)$ for all $i\neq j$ and thus $c_j\in\overline D_{i}$ if $j<i$. However, $c_j\notin\dom(u_j)$ and hence $c_j\notin\dom_i$ if $j\ge i$. For~\ref{itm:2ii}, merely observe that $c_i$ is the Condorcet winner in $\dom_{{i+1}}(c_j)$, if $j<i\le n$.  To appreciate~\ref{itm:2iii}, observe that by construction $\dom_{{i+1}}(c_i)$ has to be either a singleton $\set{u_i}$ for some~$u_i\in U_i$, or $U_i$ itself.  The former is the case if $U_i\subseteq Y$, or if $U_i\subseteq X$ and $i\neq n$.  The latter holds if $U_i=U_{n}$ or if $U_i\subseteq Z$.  In either case, $\teqset[\dom_{i+1}(c_i)]=\dom_{i+1}(c_i)$ and $u_i\teqrel[i+1]c_i$ holds. For the case in which $U_i\subseteq X$ with $i\neq n$, let~$U_i=\set{u_i,u_i',u_i''}$. By construction, $U_{i+2}\subseteq Z$ and $u_i',u_i''\notin\dom(u_{i+2})$. Accordingly, $u_i',u_i''\notin\dom_{i+1}$. Form $u_i\in\dom_{i+1}$ it then follows that $\dom_{i+1}\cap U_i=\set{u_i}$.
	
	We are now in a position to prove~(\ref{eqn:dinteq}) by induction on~$k$.  For $k=1$, observe that~$d$ is a Condorcet winner in~$\overline D_1$ and thus $d\in\teqset[\overline D_1]$. For the induction step, let $k=i+1$. With observation~\ref{itm:2i} we know that $u_i\in \overline D_{i+1}$ and, in virtue of the induction hypothesis, also that $d\in \teqset[{\overline D_i}]$. Hence, $d\teqrel[i+1]u_i$. Moreover, by observations~\ref{itm:2ii} and~\ref{itm:2iii}, 
	$c_i\teqrel[i+1]d\teqrel[i+1] u_i\teqrel[i+1] c_i$, \ie 
	$c_i$, $d$ and $u_i$ constitute a $\teqrel[i+1]$-cycle. In virtue of Lemma~\ref{lemma:CteqdomX} and observation~\ref{itm:2half}, we may conclude that $c_i\teqrel[i+1]^*a$ for all $a\in\dom_{i+1}$.
	Accordingly, $\set{c_i,d,u_i}\subseteq\tc_{\dom_{i+1}}(\teqrel[i+1])$ and $d\in\teqset[\overline D_{i+1}]$, which concludes the proof.\qed
	\end{proof}
	
	\bibliography{../../bib/abb,../../bib/brandt,../../bib/fischerf,../../bib/pamas}

\end{document}

%% file: Tournament_Example.tex
\begin{tikzpicture}[scale=.3]
  \tikzstyle{every circle node}=[draw,minimum size=1.6em,inner sep=0pt]
	\draw (0,0) 	node[circle](a){$b$}	++(8,0) node[circle](b){$c$}	++(-4,5.5) node[circle](c){$a$}	++(-4,-12)	node[circle](d){$d$}	++(8,0) node[circle](e){$e$};
	\draw[-latex, very thick] (c)--(a);
	\draw[-latex, very thick] (a)--(b);
	\draw[-latex, very thick] (b)--(c);
	\draw[-latex] (a)--(d);
	\draw[-latex] (e)--(a);
	\draw[-latex, very thick] (c) .. controls (13,4) and (14,-5) .. (e);	
	\draw[-latex, very thick] (c) .. controls (-5,4) and (-6,-5) .. (d);
	\draw[-latex, very thick] (d)--(e);
	\draw[-latex, very thick] (b)--(e);
	\draw[-latex] (d)--(b);
\end{tikzpicture} 

%% file: Banks_construction_form.tex
\begin{tikzpicture}[scale=.8]
  \tikzstyle{every ellipse node}=[draw,inner xsep=4em,inner ysep=1em,fill=black!15!white,   draw=black!15!white
  ]
  \foreach \x / \y in {4/2,8/1,6/1a}
	\draw  (1.5,\x) node[ellipse] (ellipse\y){};
  \tikzstyle{every circle node}=[draw,minimum size=1.6em,inner sep=0pt]
  	\draw (0,4)		node[circle](x21){$p$} ++(1.5,0) node[circle](x22){$q$} ++(1.5,0) node[circle](x23){$\neg r$};
	\draw (0,6)		node[circle](z11){$p$} ++(1.5,0) node[circle](z12){$s$} ++(1.5,0) node[circle](z13){$r$};
  	\draw (0,8)		node[circle](x11){$\neg p$} ++(1.5,0) node[circle](x12){$s$} ++(1.5,0) node[circle](x13){$q$};
  	\draw (-2,5)	node[circle](y2){$y_2$};	
	\draw (-2,7)	node[circle](y1){$y_1$};
	\draw (-1,9.75)		node[circle](d){$d$};
	\draw (0,11)		node[circle](c1){$c_1$} ++(0,1.5) node[circle](c1a){$c_3$} ++(0,1.5) node[circle](c2){$c_5$};
	\draw (-2,11.75)		node[circle](c1b){$c_2$} ++(0,1.5) node[circle](c2b){$c_4$};
	\foreach \x in {1,2}
	 \draw[-latex] (x\x3) .. controls ++(-1,.7) and ++(1,.7) .. (x\x1);
	\foreach \x in {1,2}
	 \draw[-latex] (x\x1) -- (x\x2);
	\foreach \x in {1,2}
	 \draw[-latex] (x\x2) -- (x\x3);
    \foreach \x in {1} \foreach \y / \z in {1/2,2/3}
	 \draw[-latex] (z\x\y) -- (z\x\z);
    \foreach \x in {1}
	 \draw[-latex] (z\x3) .. controls ++(-1,.7) and ++(1,.7) .. (z\x1);
	\draw[-latex] (y1) .. controls  +(-3.5,0) and +(-2.5,0) .. (c1b);
	\draw[-latex] (y2) .. controls +(-5.25,0) and +(-4,0) .. (c2b);
	\draw[-latex] (ellipse1) .. controls ++(3.5,0) and +(5.5,0) .. (c1);
	\draw[-latex] (ellipse2) .. controls ++(5,0) and +(7,0) .. (c2);
    \draw[-latex] (ellipse1a) .. controls ++(4.25,0) and +(6.25,0) .. (c1a);
	\draw[-latex] (x21) .. controls ++(-1,1) and ++(-1,-1) .. (x11);
    \draw[-latex] (x23) -- (z13);
    \draw[-latex] (z11) -- (x11);
\end{tikzpicture} 

%% file: TEQ_construction_form.tex
\begin{tikzpicture}[scale=.8]
  \tikzstyle{every ellipse node}=[draw,inner xsep=4em,inner ysep=1em,fill=black!15!white,   draw=black!15!white
  ]
  \foreach \x / \y in {0/3,4/2,8/1,2/2a,6/1a}
	\draw  (1.5,\x) node[ellipse] (ellipse\y){};
  \tikzstyle{every circle node}=[draw,minimum size=1.6em,inner sep=0pt]
  	\draw (0,0)  		node[circle](x31){$p$} ++(1.5,0) node[circle](x32){$q$} ++(1.5,0) node[circle](x33){$\neg r$};
	\draw (0,2)		node[circle](z21){$z_2^1$} ++(1.5,0) node[circle](z22){$z_2^2$} ++(1.5,0) node[circle](z23){$z_2^3$};
  	\draw (0,4)		node[circle](x21){$p$} ++(1.5,0) node[circle](x22){$s$} ++(1.5,0) node[circle](x23){$r$};
	\draw (0,6)		node[circle](z11){$z_1^1$} ++(1.5,0) node[circle](z12){$z_1^2$} ++(1.5,0) node[circle](z13){$z_1^3$};
  	\draw (0,8)		node[circle](x11){$\neg p$} ++(1.5,0) node[circle](x12){$s$} ++(1.5,0) node[circle](x13){$q$};
 	\draw (-2,1)  	node[circle](y4){$y_4$};
	\draw (-2,3)	node[circle](y3){$y_3$};
  	\draw (-2,5)	node[circle](y2){$y_2$};	
	\draw (-2,7)	node[circle](y1){$y_1$};
	\draw (-1,9.75)		node[circle](d){$d$};
	\draw (0,11)		node[circle](c1){$c_1$} ++(0,1.5) node[circle](c1a){$c_3$} ++(0,1.5) node[circle](c2){$c_5$} ++(0,1.5) node[circle](c2a){$c_7$} ++(0,1.5) node[circle](c3){$c_9$};
	\draw (-2,11.75)		node[circle](c1b){$c_2$} ++(0,1.5) node[circle](c2b){$c_4$} ++(0,1.5) node[circle](c3b){$c_6$} ++(0,1.5) node[circle](c4b){$c_8$};
	\foreach \x in {1,2,3}
	 \draw[-latex] (x\x3) .. controls ++(-1,.7) and ++(1,.7) .. (x\x1);
	\foreach \x in {1,2,3}
	 \draw[-latex] (x\x1) -- (x\x2);
	\foreach \x in {1,2,3}
	 \draw[-latex] (x\x2) -- (x\x3);
    \foreach \x in {1,2} \foreach \y / \z in {1/2,2/3}
	 \draw[-latex] (z\x\y) -- (z\x\z);
    \foreach \x in {1,2}
	 \draw[-latex] (z\x3) .. controls ++(-1,-.7) and ++(1,-.7) .. (z\x1);
	\draw[-latex] (y1) .. controls  +(-4.5,0) and +(-3.5,0) .. (c1b);
	\draw[-latex] (y2) .. controls +(-6.25,0) and +(-5,0) .. (c2b);
	\draw[-latex] (y3) .. controls +(-7.5,0) and +(-6.25,0) .. (c3b);
	\draw[-latex] (y4) .. controls +(-8.75,0) and +(-7.75,0) .. (c4b);
	\draw[-latex] (ellipse1) .. controls ++(3.5,0) and +(5.5,0) .. (c1);
	\draw[-latex] (ellipse2) .. controls ++(5,0) and +(7,0) .. (c2);
	\draw[-latex] (ellipse3) .. controls ++(6.5,0) and +(8.5,0) .. (c3);
    \draw[-latex] (ellipse1a) .. controls ++(4.25,0) and +(6.25,0) .. (c1a);
    \draw[-latex] (ellipse2a) .. controls ++(5.75,0) and +(7.75,0) .. (c2a);
	\foreach \x in {1,2} \foreach \y / \z in {1/2,1/3,2/1,2/3,3/1,3/2}
    \draw[-latex] (z\x\y) -- (x\x\z);
	\draw[-latex] (x21) .. controls ++(-1,1) and ++(-1,-1) .. (x11);
	\draw[-latex] (x33) .. controls ++(1,1) and ++(1,-1) .. (x23);
	\draw[-latex] (x31) .. controls ++(-1.75,0.75) and ++(-1.75,-0.75) .. (x11);
\end{tikzpicture} 